\documentclass[a4paper,fleqn,usenatbib]{mnras}
\usepackage{amssymb,amsmath,graphicx,psfrag,mathtools}
\usepackage[T1]{fontenc} 
\usepackage{ae,aecompl} 
\usepackage{url}
\title[Fast Radio Burst Energetics and Sources]{Fast Radio Burst Energetics and Sources}
\author[J. I. Katz]{
J. I. Katz,$^{1}$\thanks{E-mail katz@wuphys.wustl.edu} 
\\
$^{1}$Department of Physics and McDonnell Center for the Space Sciences,
Washington University, St. Louis, Mo. 63130 USA 
}
\date{Accepted XXX.  Received YYY; in original form ZZZ} 
\pubyear{2019} 
\date{\today}
\begin{document} 
\label{firstpage} 
\pagerange{\pageref{firstpage}--\pageref{lastpage}} 
\maketitle 
\begin{abstract}
	{ Repeating and apparently non-repeating fast radio bursts (FRB)
	differ by orders of magnitude in duty factors, energy and rotation
	measure.  Extensive monitoring of apparently non-repeating FRB has
	failed to find any repetitions.  This suggests the two types differ
	qualitatively, rather than in repetition rate, and are produced by
	distinct kinds of sources.} The absence of periodicity in repeating
	FRB argues that they are not produced by neutron stars.  They may
	originate in { dilute} relativistic jets { produced by low
	luminosity black hole accretion.  Non-repeating FRB may be produced
	by} catastrophic events such as the collapse of an accreting
	magnetic neutron star to a black hole { or of an accreting
	magnetic white dwarf to a neutron star, during which} a disappearing
	magnetic moment { radiates dipole radiation that accelerates
	electrons in nearby matter.}  If they are emitted by collimated
	beams of relativistic particles or charge ``bunches'' with Lorentz
	factor $\gamma$, their radiation is collimated into a solid angle
	$\sim \gamma^{-2}$ sterad, reducing the energy requirement.  If
	powered by magnetic reconnection, a pulse of length $\Delta t$ may
	draw on the magnetic energy in a volume $\sim \gamma^4
	(c \Delta t)^3$, although only a fraction $\sim 1/\gamma^2$ of this
	may be dissipated without decollimation.
\end{abstract}
\section{Introduction}
{ Fast Radio Bursts (FRB) are rare events.  The rate of FRB with fluences
at $\sim 1400$ MHz $\gtrsim 1$ Jy-ms (roughly the Parkes detection
threshold) is controversial, but estimates are within an order of magnitude
of $10^6$/sky-y \citep{L18}.  Combining with an estimate \citep{C16} of
$\sim 10^{11}$ $L^*$ galaxies within $z \lesssim 1$ implies a rate of $\sim
10^{-5}$/galaxy-y, three or four orders of magnitude less than the supernova
rate \citep{L11}. FRB require rare progenitors.  If all FRB repeat their
sources must be even rarer than the bursts themselves, and if only the
fraction known to repeat actually do so, the repeaters must be rarer still.

FRB do not have obvious identifications with other astronomical objects
(\citet{T17} identified FRB 121102 with an unremarkable dwarf galaxy).
Their sources must be rare and special; if they were abundant, scaling with
the density of stars, the luminosity of an active galactic nucleus (AGN), or
some other observationally obvious parameter, those at comparatively low
redshift would be expected (even with poor localization) to be identified
with rich clusters of galaxies, luminous AGN, or other prominent hosts.
This gives the modeler great freedom to consider exotic circumstances but
makes it difficult to exclude hypotheses on the basis of the special
conditions required.}

If FRB radiate isotropically, the cosmological distances implied by
attributing their dispersion measures to intergalactic plasma in standard
cosmology imply instantaneous { isotropic equivalent} luminosities as
high as $\sim 10^{43}$ ergs/s and energies $\sim 10^{40}$ ergs \citep{T13}.
The identification of the { first} repeating FRB 121102
\citep{C17,M17,T17,B17} with a dwarf galaxy at redshift $z = 0.193$ { and
the second at $z \le 0.1$ \citep{CHIME1}} implies their burst energies are a
few orders of magnitude lower.

FRB must be emitted in low density plasma because $\sim 1\,$GHz radiation
cannot propagate through plasma with $n_e \gtrsim 10^{10}$ cm$^{-3}$.  In
the plasma that produces most of the dispersion, likely some combination of
interstellar media in a host galaxy and our Galaxy and the intergalactic
medium, $n_e < 5 \times 10^7$ cm$^{-3}$ because in the relation $\Delta t
\propto \nu^\alpha$ the dispersion index $|-\alpha -2| \le 0.003$
\citep{K16a}.  { Higher $n_e$ is possible in regions that are only small
contributors to the dispersion.}

Do repeating and non-repeating FRB have the same sources and emission
mechanisms, differing only in repetition rate, or are they fundamentally
different, as suggested by \citet{CSS18,PLZ18}?  Must a single model be
found, that with only quantitative modification can explain both repeating
and non-repeating FRB, or could the prime movers of these classes of events
differ qualitatively?  Can qualitatively different prime movers produce
similar radiation, perhaps by the same physical processes?  Analogous
questions arose in the study of gamma-ray bursts (GRB), that could not be
understood until it was recognized that soft gamma repeaters (SGR) { are
a different phenomenon} \citep{K02,Ku18}.

Relaxing the requirement that a single model account both for non-repeating
FRB and for a repeating FRB that has been observed over five years frees the
modeler from constraints that may be mutually contradictory.  There is much
more freedom in modeling non-repeating FRB if they can be catastrophic
events, destroying the source objects, { and in modeling repeating FRB if
they are not required to meet the energetic requirements of the
non-repeaters}.

Several differences between the repeating and non-repeating FRB point to
their being distinct phenomena:
\begin{enumerate}
	\item The difference between repetition and non-repetition is
		quantified by the duty factor
\begin{equation}
	\label{dutyfactor}
	D \equiv {\langle {\cal S} \rangle^2 \over
	\langle {\cal S}^2 \rangle}
\end{equation}
\citep{K16b}, where $\cal S$ is the flux density.  The repeater FRB 121102
had $D \sim 2 \times 10^{-6}$ during a period of high activity \citep{Z18},
while for non-repeating FRB $D < 10^{-9}$ may be inferred from their
long-term absence of repetitions \citep{S18}\footnote{These values depend on
the sensitivity of the observations and can only be lower limits because any
emission below the detection threshold is not included.  See Sec.~\ref{duty}
for discussion.}.  This difference of at least three orders of magnitude in
$D$ suggests that the sources are qualitatively different.
\item The instantaneous { isotropic-equivalent} power implied by the
	repeating bursts is { about three orders of magnitude} less than
		that required by the non-repeaters because of the repeaters'
		lower intensity and smaller distances:
		$z = 0.193,\ \le 0.1$ \citep{T17,CHIME1} rather than $z \sim
		1$ { for many non-repeaters}.  In addition, repeaters may
		be collimated \citep{K17a}, { further reducing their
		required power}, while in some models the event rate of the
		non-repeaters implies that they are not collimated
		(Sec.~\ref{rate}).
%
	\item { The rotation measure (RM) of the repeating FRB 121102 (it
		has not been reported for the second repeater FRB
		180814.J0422+73 \citet{CHIME1}) is $\gtrsim 10^3\ \times$
		the measured RM of non-repeating FRB \citep{G18,Mi18}.}
		This also suggests a qualitative difference.
\item { Both repeaters show downward spectral drifts within their pulses
	\citep{G18,Mi18,H19,CHIME1}; this has not been reported in
	non-repeaters.}
\item { Non-repeating FRB have a characteristic energy scale while the
	bursts of the repeating FRB 121102 appear not to have such a scale.}
\end{enumerate}

{ Sec.~\ref{RM} discusses the differences in rotation measure.  They
imply that repeating and non-repeating FRB originate in different
environments, and therefore likely are of different nature.}
Sec.~\ref{ns} considers the hypothesis of neutron star origin and discusses
its weaknesses.  Sec.~\ref{periodicity} discusses the failure to detect
periodicity in the repeating FRB 121102, an argument against neutron star
models.  Sec.~\ref{sky} discusses the distribution of non-repeating FRB on
the sky.  This implies that their sources are rare and that there is a
characteristic strength of their outbursts; none occurred in the surveyed
region of the Galaxy (no { low energy} Galactic ``nano-FRB'' have been
detected). 

The problem of describing the emitting region and mechanism should be
separated from the problem of the prime mover, the ultimate energy source.
Non-repeating and repeating FRB may have similar emission processes but
different prime movers.  Sec.~\ref{reconnection} explores the energetics of
such a possible physical process: magnetic reconnection of two discrete low
density regions of uniform (on the scale on which magnetic reconnection
takes place) magnetic induction.  { Magnetic reconnection} can produce
collimated radiation { if it accelerates relativistic particles}.
Sec.~\ref{energetics} applies this to FRB and pulsar nanoshots.

Sec.~\ref{repsites} presents a model of repeating FRB resulting from
magnetic reconnection in long-lived regions of turbulent gas flow, such as
relativistic jets produced by accretion onto black holes (accretion discs
themselves are too dense to permit the escape of GHz radiation).  Repetition
may continue indefinitely.  There is no evident characteristic scale of
outburst, and the sources, AGN or { other accreting massive black holes},
are rare.  Such a model avoids the energetic problems of neutron star models
of the repeating FRB 121102, especially those that describe FRB as scaled-up
giant pulsar pulses \citep{K17a,W18}.

Sec.~\ref{nonrep} discusses the origin of non-repetitive FRB in catastrophic
events that destroy their sources.  Even if not ``standard flashbulbs'',
their total emitted energy appears to be dominated by the most energetic
outbursts.  These events may be the collapse of accreting magnetic neutron
stars to black holes, { or of accreting magnetic white dwarfs to neutron
stars}.

{ Conclusions are summarized in Sec.~\ref{conclusions}.}
\section{Rotation Measure}
\label{RM}
{ Rotation Measure, defined by the integral
\begin{equation}
	\text{RM} \equiv \int\! n_e B_\parallel\,d\ell
\end{equation}
along the line of sight (and usually parametrized by the rotation angle
of linear polarization $\Delta\theta(\lambda)/\lambda^2$), { is dominated
by} the near-source environment.  Although the intergalactic medium may be
the dominant contributor to dispersion measure (DM), it contributes
negligibly to RM.  Both electron density and magnetic field are small in the
intergalactic medium, compared to their interstellar or near-source values,
and RM is second-order in these small quantities.}

Rotation Measures may be significant clues to the origins of FRB.  The
repeating FRB 121102 has $\text{RM} \approx 1.4 \times 10^5$ rad/m$^2$,
which varied by about 10\% in seven months \citep{Mi18}.  This is
reminiscent of PSR J1745-2900 with $\text{RM} \approx 7 \times 10^4$
rad/m$^2$, varying 5\% in 1--2 years \citep{D18}.  PSR J1745-2900 has a
projected separation of 0.097 pc from the Galactic Center black hole source
Sgr A$^*$ \citep{B15}.  { Sgr A$^*$ itself} has $\text{RM} \approx -5 \times
10^5$ rad/m$^2$ \citep{BBD18}.

Analogy suggests that FRB 121102 is causally associated with or even
identical to the persistent radio source in its host galaxy.  This
persistent source is consistent with being a low luminosity AGN \citep{M17}.
The measured upper limit on their projected separation of 40 pc, $\sim 1$\%
of the size of the galaxy, is small enough to render an accidental
coincidence implausible.

The { measured RM} of non-repeating FRB { are all} smaller by { at
least three} orders of magnitude { than that of FRB 121102 (the RM of the
second repeater FRB 180814.J0422+73 has not been measured \citet{CHIME1})},
and are comparable to those of Galactic pulsars \citep{Mi18}.  The most
extreme example is FRB 150215, for which $-9 < \text{RM} < 12$ rad/m$^2$
\citep{P17}.  This indicates that { non-repeaters are distinct from
repeating FRB, occur in different environments and therefore likely have
different sources.}  Non-repeaters are not associated with galactic nuclei,
{ that have much higher RM \citep{P16}.}
\section{Why Not Neutron Stars}
\label{ns}
Many models of FRB { (see reviews by \citet{K18a,P18})} assume their
sources are magnetic neutron stars and their origin is either as super-giant
pulsar pulses or in SGR outbursts.  Magnetic neutron stars are candidates
because their small size and high magnetic, gravitational and rotational
energy densities appear a natural match to the brevity and high power of FRB.
The facts that both radio pulsars and FRB have extraordinary brightness
temperatures, implying coherent emission, and that some pulsars emit
occasional giant pulses, have made giant pulsar (PSR) pulses writ large a
popular model of FRB.
\subsection{Pulsar Models}
The super-giant pulsar pulse model of FRB has the difficulty that in
standard pulsar models the instantaneous radiated power cannot exceed the
instantaneous spindown power.  The requirement of an instantaneous power as
large as $10^{43}$ ergs/s, without any intermediate energy reservoir between
rotational energy and radiation, would demand extreme values of both the
pulsar's magnetic moment and its rotation rate.  { Appendix
\ref{origspin} also argues that neutron stars are not born with the required
combination of fast (near-breakup) spin and huge (``magnetar'') magnetic
moment.}

The rotational energy store of a neutron star is limited by its equation of
state (and hence its maximum rotation rate), independent of its magnetic
moment.  In a pulsar model that store must be no less than the {\it peak\/}
(burst) radiated power multiplied by the active lifetime of the neutron star
as a source of FRB.  These values are so extreme as to verge on
contradicting the lower bound on { (repeating)} FRB lifetime set by the
observation of FRB 121102 over six years \citep{K18a}.  This problem might
be avoided if there is an energy store intermediate between rotation and
bursts that can be drawn upon for short periods, as a discharging capacitor
\citep{K17b} { powers a spark}.

{ Collimation \citep{K17a} provides another possible loophole by reducing
the required power from its isotropic-equivalent value, perhaps by a large
factor.}  The extraordinary brightness temperatures of FRB require radiation
by charge ``bunches'' whose net charges are so large that their
electrostatic potentials are $\gg m_e c^2/e$, implying that they can only
hold together if the radiating electrons are highly relativistic
\citep{K14,K18b}.  Radiation by relativistic particles is collimated if the
particle velocity vectors are themselves collimated.   In a pulsar 
magnetosphere a particle moving along a magnetic field line sweeps its
radiation cone over an angle $\sim 1$ radian, and the field lines themselves
spread over $2 \pi$ radians in azimuth, { arguing against the collimation
loophole to the energy requirement.}
\subsection{Soft Gamma Repeater Models}
An alternative model, also based on a magnetic neutron star, associates FRB
with Soft Gamma Repeater (SGR) outbursts \citep{K16b,K18a}.  It has several
difficulties: 1) No FRB was observed simultaneous with the great outburst of
the Galactic SGR 1806$-$20 \citep{TKP16}; even far out-of-beam, a Galactic
FRB would be about 50 dB more intense than one at cosmological distance.
2) SGR are thermal emitters { with a dense photon-pair gas environment that
would rapidly degrade energetic particles by Compton, Bhabha and M{\o}ller
scattering and prevent the development of large accelerating electric fields
by the enormous conductivity of the pair gas}.  3) The activity of FRB
121102 is not modulated at periods characteristic of SGR \citep{Z18} (or any
other period).  
\section{Absence of Periodicity}
\label{periodicity}
Two distinct kinds of periodicity should be distinguished.  In {\it absolute
periodicity\/}, the intervals between pulses are integer multiples of an
underlying period, while in {\it continuous modulation\/} the rate of bursts
or their intensity is periodically (not necessarily sinusoidally) modulated,
but individual bursts need not be separated by integer multiples of a
period.  In the limit as the modulation function approaches a Dirac
$\delta$-function, continuous modulation becomes absolute periodicity.  Slow
continuous changes in periods may be readily included in these models.
\subsection{Absolute Periodicity}
Radio pulsars are absolutely periodic, aside from their gradual slowing
and infrequent small discontinuous period changes (glitches).  Their
pulses are observed at most or all periodically spaced times (some pulses
may be absent, or ``nulled'').  Rotating Radio Transients (RRAT) \citep{M06}
are Galactic pulsars, the overwhelming majority of whose possible pulses are
nulled.  This could be the result of no pulse being emitted or of emitted
pulses that are not directed toward the observer.  RRAT are discovered as
single pulses spaced by large multiples of the underlying period and whose
periodicity may only be revealed by sustained monitoring.

Absolute periodicity is comparatively easy to detect, or to exclude, because
even a single out-of-phase burst is sufficient to exclude that combination
of period and phase.  Approximately $\log_2{(T_{span}/P_{min})}/
(-\log_2{D})$ bursts must be observed, where $D$ is the duty factor (the
fraction of the rotational period during which a burst may be emitted,
including any measurement uncertainty) to be confident of detecting, or
excluding, a period greater than $P_{min}$ in a span of data $T_{span}$.
Very close burst pairs, if representing separate bursts rather than
substructure of longer bursts, are equivalent to small $T_{span}$ and are
powerful tools for constraining possible absolute periodicity \citep{K18c}.
\subsection{Continuous Modulation}
In an alternative model, the brightness or flux of bursts is modulated
continuously, such as by neutron star rotation.  Bursts are equally likely
to occur at any modulation phase, but will be brighter and more often
detected at some phases than at others.  Such modulation is likely in
neutron star models in which emission is not narrowly collimated, but in
which radiation is brighter when the emitting region is on the side of the
neutron star facing the observer, and not occulted by the star.  This
{ describes the modulation of anomalous X-ray pulsars (AXP/SGR) and of
accreting binary neutron stars.}

Continuous modulation is inconsistent with narrow collimation { in a
direction fixed in the frame of the rotating source}.  A narrowly
collimated rotating beam will only be observable at times separated by
integer multiples of the rotation period, when it points to the observer;
this is absolute periodicity.  Continuously modulated models therefore
require that FRB be uncollimated and extremely luminous, with actual
luminosities comparable to their isotropic-equivalent luminosities.

In continuously modulated models periodicity may be manifested as bursts at
favorable phases are more readily detected, or as bursts as those phases are
brighter, on average, than those at less favorable phases.  This second
effect occurs only if there is an intrinsic characteristic scale of burst
brightness, fluence, or other detectability parameter independent of
modulational phase.  It does not occur if the distribution of the
detectability parameter is a power law $N(L) \propto L^{-\beta}$, because
then $\langle L \rangle_{L \ge L_0} = L_0 (1-\beta)/(2-\beta)$ (the finitude
of the total emission requires $\beta > 2$ and a lower cutoff below the
range of observation, as described in Sec.~\ref{sky}).  Because bursts can
occur at any phase of the periodic modulation, an out-of-phase burst does
not exclude a period, but only renders it less likely in a Bayesian sense.

Detection or exclusion of periodic modulation
is more difficult than detection of absolute periodicity.  If the modulation
is sinusoidal with fractional amplitude $a$ it requires the detection of
roughly $\log_2{(T_{span}/P_{min})}/a$ bursts, typically several dozen.  
\subsection{FRB 121102}
Examination of the very short (3--100 ms) burst intervals reported by
\citet{S17,H17,Z18} excludes any absolute periodicity unless these short
intervals represent substructure of longer ($\gtrsim 100$ ms) bursts rather
than separate bursts.  If bursts are that long, the necessarily even longer
periods are excluded by timing of the longer intervals.  The analysis of
\citet{Z18} can also be interpreted as setting an upper bound on the
amplitude of any continuously modulated periodicity of $\lesssim 30$\%.

Pulsars are known with much shorter periods than the absolute periods $\ge
10$ ms excluded by \citet{Z18}, and energetic considerations require very
short periods in pulsar models of FRB \citep{K18a}.  However, the failure to
detect periodicity encourages consideration of models that have no
periodicity of either kind, and therefore that do not involve a rotating
neutron star.

\section{Sky Distribution and Rarity of FRB}
\label{sky}
FRB sources are rare in the Universe.  This is shown by their dispersion
measures, substantially attributed to intergalactic plasma and implying
cosmological distances.  { Statistical arguments are possible only for
the non-repeating FRB.}

If the sources of { non-repeating} FRB were distributed similarly to
stars, as expected in any model that associates them with objects related to
stars, the distribution of their flux on the sky would resemble that of
starlight.  It does not.  { Non-repeating} FRB are not concentrated in
the Galactic plane \citep{B18}.  Yet most of the starlight in the sky is
Galactic, and concentrated in the plane: At $\lambda = 2\text{--}2.5\,\mu$
(K-band) where extinction is comparatively small, the extra-Galactic
background light is $\sim 10$ nW m$^{-2}$ sr$^{-1}$ \citep{D11} while the
Galactic starlight is $\sim 300$ nW m$^{-2}$ sr$^{-1}$ at $b = 30^\circ$ and
even more in the Galactic plane \citep{L98}.  Allowing for extinction
further increases the ratio of plane to high latitude starlight and the
difference between the distributions of starlight and of { non-repeating}
FRB.  Either:
\begin{enumerate}
	\item The density of { non-repeating} FRB sources does not follow
		that of stars, or
	\item { Non-repeating} FRB sources are so rare that {\it no\/}
		Galactic outbursts have been observed.
\end{enumerate}

The first explanation would indicate that { non-repeating} FRB are not
emitted by objects associated with stars or their descendants, such as
neutron stars { (X-ray binaries and pulsars)}.  Such a model is considered
for repeating FRB in Sec.~\ref{repsites}.

The second explanation would require that { non-repeating} FRB sources,
including hypothetical sources of more frequent but extremely low energy
nano-FRB { in the Galaxy}, are also too rare { to have been detected.  It
argues against pulsar models (Sec.~\ref{ns}) that would be expected to have
many small outbursts for each FRB super-outburst.}  Averaging over
sufficient time and solid angle would lead to a flux distribution on the sky
resembling that of stars, although the required average might be over
thousands or millions of sterad-years.

The second explanation is consistent with { non-repeating} FRB sources
that are very rare neutron star events or subspecies, {\it none} of which
are now active in the Galaxy.  These might be neutron stars less than a few
decades old, or with unprecedented magnetic fields, or undergoing very rare,
{ perhaps catastrophic,} events such as { the collapse} discussed in
Sec.~\ref{nstobh}.  It would require the existence of a large characteristic
energy scale, so that the { non-repeating} FRB flux during the period and
in the directions of observation when no characteristic bursts are observed
is much less than its mean, averaged over sufficient time and solid angle.
SN, GRB and SGR have such characteristic scales, but known stellar flares
and giant pulsar pulses do not; their flux is dominated by the weakest
events \citep{CBH04,PS07,KP17,MSB18}.  The distribution of bursts from the
repeating FRB 121102 is not well determined, but within the dynamic range of
observation their fluence is dominated by weak bursts \citep{L17,Z18,G19};
{ they do not have a characteristic scale in the sense used here}.

{ There are no Galactic FRB among the 65 observed FRB \citep{frbcat}.  With
about 95\% confidence, the Galactic fraction $f_G \lesssim 0.06$.  The ratio
of weak non-repeating bursts detectable at Galactic distances (only from one
galaxy, ours) to those detectable from the $N_g \sim 10^{11}$ galaxies in
the Universe with $z \lesssim 1$ must be $< f_G N_g \approx 6 \times 10^9$.
Galactic nano-FRB would be detectable with emission $E_G \approx
(\text{10 kpc/3 Gpc})^2 \sim 10^{-11}$ of those of cosmological FRB.  If the
distribution of emission strengths is a power law $n(L) \propto L^{-\beta}$,
the exponent $\beta < -\ln{f_G N_g}/\ln{E_G} - 1 \approx - 0.1$.  Any energy
emitted by Galactic nano-FRB with the same S/N as the observed cosmological
FRB is much less than that emitted by the cosmological FRB.  This does not
contradict the expectation that, averaged over a sufficiently long time, if
FRB are produced by a stellar-related population, the FRB fluence received
from the very infrequent Galactic FRB will far exceed that of extra-Galactic
FRB because of the proximity of Galactic objects.}

If FRB have a characteristic energy scale, with their total mean flux
dominated by infrequent large outbursts, then the rarity of FRB implies the
rarity of outbursts but not the rarity of their sources, { whose
outbursts could be very infrequent or occur only once}.  For example, the
Galaxy could contain many proto-FRB that will produce a burst once in the
distant future.

The slope of the distribution of FRB fluences may be too uncertain
\citep{ME18a,ME18b,J18} to address this question, but the fact that the
high-fluence ASKAP FRB \citep{S18} have a mean DM (and hence distance) about
half that of the Parkes FRB \citep{LYB19} weakly constrains the distribution
of fluences---detections by the more sensitive Parkes telescope are not
dominated by less luminous bursts from nearby sources (Sec.~\ref{nonrep}).
This is consistent with the inference from the distribution of FRB on the
sky.  The slopes of the distributions $N(DM)$ and $N(\text{Fluence})$ of the
ASKAP FRB \cite{LP19} are similar to those of the Parkes FRB \cite{K16a}.

\section{Magnetic Reconnection}
\label{reconnection}
{ Magnetic reconnection, the rapid dissipation of magnetostatic energy
in a thin current sheet, is an incompletely understood phenomenon that is
known to accelerate charged particles in plasmas ranging from the laboratory
to the Sun, the solar wind and its interaction with planetary magnetospheres
and is hypothesized to explain many other cosmic phenomena
\citep{LAD11,GZ16}.  Like other collective fluid structures, it can occur
when microscopic dissipative processes are slow (Reynolds or magnetic
Reynolds numbers are large).  It has recently been proposed to explain {
some forms of} the coherent radio emission of pulsars \citep{PUSC19}.  Here
I suggest it as a possible mechanism for the acceleration necessary to make
coherently radiating relativistic charge ``bunches'' that emit FRB.}

Two regions of differing magnetic induction are
separated by a thin current sheet.  In a force-free configuration
(${\vec J}\times{\vec B} = 0$) the electrons move along the magnetic field
lines.  If the magnetic induction is nearly unidirectional through the
reconnection region, the total radiation has the relativistic collimation of
the radiation by individual particles (or charge ``bunches'').  However, a
current ${\vec J} \parallel {\vec B}$ leads to a rotation by an angle
$\Delta \theta$ of the direction of $\vec B$ through the current sheet while
its magnitude does not change.  To
maintain the collimation of relativistic motion with Lorentz factor $\gamma$
requires $\Delta \theta \lesssim 1/\gamma$, and $\vert\Delta {\vec B}\vert
\lesssim \Delta \theta \vert{\vec B}\vert \sim \vert{\vec B}\vert/\gamma$.

As a result, only a small fraction of the magnetostatic energy can be
transformed to kinetic energy of collimated relativistic charged particles.
The mean energy density that can be released by magnetic reconnection
between equal volumes with inductions $B{\hat x}$ and $B({\hat x}
\cos{\theta} + {\hat y}\sin{\theta})$ for $\theta \sim 1/\gamma \ll 1$ is
\begin{equation}
\label{calE}
{\cal E} \approx {B^2 \over 32\pi}\theta^2 \sim {B^2 \over 32\pi \gamma^2}.
\end{equation}
The inefficiency of tapping the magnetostatic energy is compensated by the
fact that the emission is narrowly collimated; the radiated power required to
provide the fluence illuminating an observer fortunate enough to be within
an angle $\le 1/\gamma$ of the beam is $\propto \gamma^{-2}$.  Of course, to
maintain the observed FRB rate and total energy radiated as FRB, { if they
are collimated} there must be a greater rate of FRB-radiating events, but as
we have no information about this rate (as opposed to the rate of those
directed toward us, that we detect), there is no inconsistency.


If magnetic reconnection is the source of FRB emission, we can set lower
bounds on the magnitudes of the fields.  Particles distributed over a length
$\Delta \ell$ in the local rest frame moving towards the observer with a
speed $v$ produce a pulse of duration
\begin{equation}
\label{deltat}
\Delta t = {\Delta\ell \over v} - {\Delta\ell \over c} \approx {\Delta\ell
\over 2 c \gamma^2}.
\end{equation}
This result yields only an upper bound on $\Delta\ell$ because $\Delta t$
may be increased by causes other than propagation delays over the path to
the observer in the emitting region.  For example, particle acceleration may
be continuous, or repeated, in a smaller radiating region.

We do not know the thickness of the region whose magnetostatic energy
contributes to the radiation received in the time $\Delta t$, but if
reconnection is rapid it may be $\sim c \Delta t$.  Because the source is
very distant from the observer the dimension in the third direction of the
radiating region is not limited by causality.  The third dimension of the
reconnecting patch is unknown, but is plausibly comparable to its
line-of-sight dimension $\Delta \ell$.  Then the radiation can draw on the
magnetostatic energy in a volume
\begin{equation}
	\label{volume}
	{\cal V} \lesssim 4 \gamma^4 (c \Delta t)^3.
\end{equation}
Because $\Delta\ell$ is an upper limit this is (aside from the assumed
size in the third dimension) an upper limit on $\cal V$.


The energy radiated over a time $\Delta t$ into a solid angle
$\sim 1/\gamma^2$ to produce an observed flux density $\cal S$ over the
bandwidth $\Delta \nu$ is
\begin{equation}
	\label{E}
	E = {{\cal S}\,\Delta \nu\,\Delta t\,d^2 \over \gamma^2},
\end{equation}
where $d$ is the distance to the FRB.  Equate this energy to that available
by magnetic reconnection in the source region
\begin{equation}
	\label{Etot}
	E = {\cal E}{\cal V}.
\end{equation}
Equations \ref{calE}, \ref{volume}, \ref{E}, and \ref{Etot} yield
\begin{equation}
	\label{B}
B \gtrsim \sqrt{8 \pi{{\cal S}\,\Delta \nu\,d^2 \over \gamma^4
c^3(\Delta t)^2}},
\end{equation}
Because Eq.~\ref{volume} is only an upper limit this result is a lower limit
on $B$ as a function of $\gamma$.  Higher $\gamma$ relax the constraints on
$B$.

{ This constraint applies to fields produced by currents in the
magnetized region.  Magnetostatic energy attributable to currents elsewhere
(for example, in a pulsar magnetosphere whose fields are largely
attributable to currents within the neutron star) cannot be dissipated by
reconnection.}

In order to produce collimated radiation { efficiently}, the beam of
``bunches'' of charge $Q$ must lose much of its energy before it is
significantly deflected by the acceleration that makes it radiate.  The
radiation time
\begin{equation}
	t_{rad} \lesssim {\rho \over \gamma c},
\end{equation}
where $\rho$ is the radius of curvature of the bunches' path.  Comparing the
energy lost in $t_{rad}$ to the kinetic energy of the electrons,
\begin{equation}
	{2 \over 3} \gamma^4 {Q^2 c \over \rho^2} t_{rad} \lesssim
	\gamma {Q \over e} m_e c^2,
\end{equation}
or
\begin{equation}
	\gamma^2 Q \gtrsim \rho {m_e c^2 \over e}.
\end{equation}
Hence if the radiation is to be collimated to the maximum degree
permitted by the Lorentz factor, that Lorentz factor must satisfy
\begin{equation}
	\label{collimLorentz}
	\gamma \gtrsim 4 \times 10^4 \left({Q \over \text{1 Coulomb}}
	\right)^{-1/2} \rho_6^{1/2}.
\end{equation}
FRB brightnesses require $Q \gg 1$ Coulomb \citep{K18b} so this condition,
although demanding, may not be impossible.  This degree of collimation is
not required by observation, so that Eq.~\ref{collimLorentz} only defines a
characteristic $\gamma$.  { Less collimated radiation is possible, but
would require greater total power, a difficulty for energy-limited pulsar
models.}

The observation of circular polarization in { one} FRB \citep{P15} ({
there are marginal detections in a few others \citep{frbcat}}) is an
additional argument in favor of radiation { by collimated beams of
charges}.  Radiation by relativistic charges moving in a plane (defined by
their instantaneous velocity and acceleration vectors) is linearly polarized
in directions in the plane of motion, but partly circularly polarized in
opposite senses on opposite sides of that plane \citep{KLB17}.  If the
radiation pattern is broadened to an angle $\gg 1/\gamma$ then the ray to
the observer will be on opposite sides of (``above'' and ``below'') planes
of motion of different radiating charges and the circular polarization will
average to a small value.  Only if the radiating particles are collimated to
an angle $\lesssim 1/\gamma$ will the ray to an observer (within an angle
$\sim 1/\gamma$ but not $\ll 1/\gamma$ of the mean plane of motion) be on
the same side of most or all of the particles' planes of motion, and a
substantial net circular polarization be observed.  However, Sec.~\ref{rate}
argues that non-repeating FRB are not collimated.
\section{Energetics}
\label{energetics}
\subsection{Poynting flux}
Jets { from accretion onto massive black holes} may be powered by
Poynting flux whose source is in the inner accretion disc.  Poynting flux
may be deposited in the boundary between a jet and a much denser accretion
{ disc or funnel, accelerating particles} that may radiate along the jet.
The power per unit area $B_{Poynting}^2 c\cos\theta/8\pi$, where $\theta$ is
the angle of incidence, may exceed that of reconnection of a quasi-static
field $\sim B_{Poynting}$ because the entire Poynting flux incident on
denser plasma may be converted to collimated radiation (rather than only the
fraction $1/\gamma^2$) if collimation is maintained by a quasi-static field.
In addition, Poynting flux propagates at the speed $c$,  while the speed of
magnetic reconnection may be much less than $c$.
\subsection{FRB}
Curvature radiation at frequency $\nu$ { by electrons accelerated by
magnetic reconnection implies $\gamma \gtrsim \gamma_{min} \equiv
(\rho \nu/c)^{1/3}$ and (Eq.~\ref{B})}
\begin{equation}
	B \gtrsim \sqrt{8 \pi {\cal S} \Delta \nu d^2 \over
	c^3 (\Delta t)^2} \left({\rho \nu \over c}\right)^{-2/3}
\end{equation}
if $\gamma \sim \gamma_{min}$.  For representative FRB parameters (${\cal S}
= 1\,\text{Jy} = 10^{-23}$ erg cm$^{-2}$s$^{-1}$Hz$^{-1}$, $d = 3\,
\text{Gpc} \approx 10^{28}$ cm, $\Delta t = 10^{-3}$ s, $\nu = 10^9$ Hz and
$\Delta \nu = 10^9$ Hz)
\begin{equation}
	B \gtrsim 9 \times 10^5 \rho_6^{-2/3} (\gamma/\gamma_{min})^{-2},
\end{equation}
where $\rho_6 = \rho/10^6$ cm.

Because of the weak ($\propto \nu^{1/3}$) dependence of the radiated power
at frequencies below the spectral peak, significant radiation may be
produced by electrons with $\gamma \gg \gamma_{min}$.  Lower values of $B$
are possible.  { The spectrum could also extend to frequencies much
greater than the $\sim 10^9$ Hz assumed in the preceding paragraph.}
%
\subsection{Pulsar nanoshots}
{ Pulsar nanoshots \citep{S04,HE07} have long presented the paradox that
these extremely brief (nanosecond) bursts of coherent radiation (that might
be considered a model of FRB, despite durations six orders of magnitude
shorter and isotropic-equivalent powers six orders of magnitude less) appear
to emit more energy than the magnetostatic energy stored in a volume $\sim
(c\Delta t)^3$.  Allowing for the relativistic effects of
Sec.~\ref{reconnection} resolves the paradox { as collimation would
reduce the required energy and magnetostatic energy could be drawn from a
larger volume.}}

Using plausible parameters
for PSR B1937+21 \citep{S04} yields (Eq.~\ref{B}) $B \gtrsim 5 \times 10^8/
\gamma^2$ gauss, and for the Crab pulsar \citep{HE07} $B \gtrsim 1.5 \times
10^8/\gamma^2$ gauss.  For $\gamma > 30$ (roughly the value required for
curvature radiation near the surface to radiate at $\sim 1\,$GHz; near the
light cylinder radius of the Crab pulsar it would imply larger $\gamma$)
these values are smaller than the dipole fields at the light cylinder radii,
which are $\sim 10^6$ gauss for each pulsar \citep{MSB18}.  The reconnection
hypothesis does not require, on energetic grounds, larger magnetic fields
than are present.

\citet{S04} used Eq.~\ref{volume}, omitting the factor of $\gamma^4$ and
assuming isotropic radiation, thus neglecting another possible factor of
$\gamma^2$, to compare the energy emitted in a nanoshot of PSR B1937+21 to
the pulsar's magnetostatic energy density, and found the disturbing result
that the emitted energy exceeded the magnetostatic energy of the source
region.  Allowing for the factors of $\gamma$ resolves this paradox.
\section{Repeating FRB}
\label{repsites}
\subsection{Accreting Black Hole Jets?} 
Massive black holes and their accretion discs and jets
\citep{RdVV16,VRBR17,K17c,Z17} may be the sites of repeating FRB.  Their
emission would not be periodic.  They would have indefinite lifetimes,
unlike pulsar-like sources whose lifetimes would be limited by spindown {
and, if long, by magnetic field decay}.  Like the accretion discs and jets
observed in X-ray binaries and AGN, they would fluctuate, with temporal
structure on a broad range of time scales.  All these properties are
consistent with observations of the repeating FRB 121102 \citep{Z18}.

\begin{figure}
	\centering
	\includegraphics[width=0.99\columnwidth]{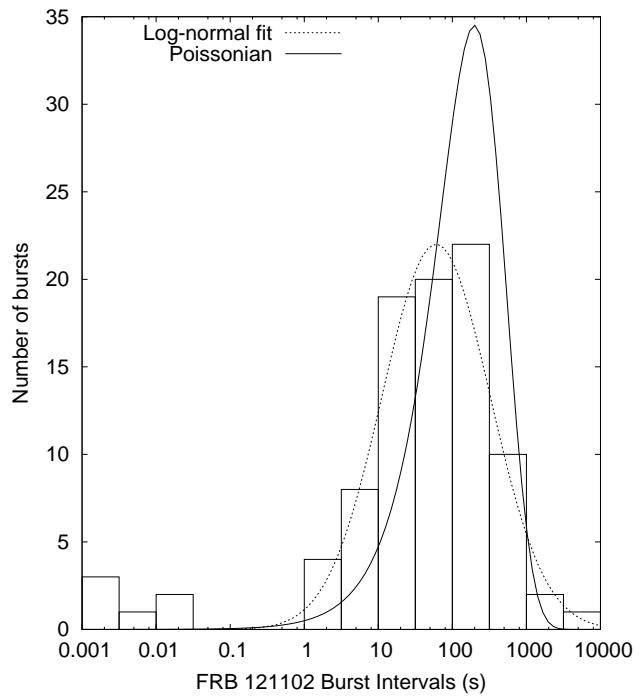}
	\caption{\label{intervals} Distribution of the 92 intervals between
	successive bursts of FRB 121102 in a five hour observation by
	Breakthrough Listen at the Green Bank Telescope at 4--8 GHz with a
	fluence threshold of about 20 Jy-$\mu$s \citep{Z18}.  A lower
	threshold would lead to shorter intervals but it is not possible to
	predict how much shorter.  The intervals $< 0.1$ s appear to be
	drawn from a different distribution than the longer intervals, and
	may represent substructure of longer bursts.  The Poissonian
	distribution assumes the observed mean burst rate of 0.005/s.  The
	log-normal function, peaking around 60 s, is a better fit and
	consistent (if the very short intervals are excluded) with the data,
	but its physical significance is unclear.  The excess of intervals
	1--30 s (compared to Poissonian) indicates correlated periods of
	activity, consistent with the existence of active and inactive
	periods on longer time scales.  { \citet{G19} reported 41 bursts
	at 1.4 GHz and found a similar distribution, log-normal with short
	interval outliers, but with a peak around 200 s.}}
\end{figure}

{ The repeating FRB 121102 has periods of greater and lesser activity
\citep{Sp16,Sc16,S17,H17,L17,G18,HSV19}.}  The statistics of the intervals
between the repetitions during one five-hour observing session { with a low
detection threshold and during which it was particularly active} \citep{Z18}
are shown in Fig.~\ref{intervals}.

The identification of FRB 121102 with { a star forming} region of a {
dwarf} galaxy \citep{M17} that also contains a persistent radio source
consistent with a weak { accreting massive black hole} \citep{C17}
supports this association: the FRB may be produced in the environment of a
black hole, perhaps by magnetic reconnection in a relativistic jet or by
deposition of Poynting flux.  The plasma frequency of such a jet, unlike
that in an accretion disc, can be low enough to permit the radiation to
escape.

The projected separation of $<40$ pc between FRB 121102 and the persistent
source would be very unlikely ($p \lesssim 10^{-3}$) if the FRB source
were drawn from a population with the same distribution as the galaxy's
visible light, spread over several kpc, { implying association with the
persistent source itself.  As discussed in Sec.~\ref{RM}, the high and
varying} rotation measure of FRB 121102 is also consistent with what we know
of the environment of the Galactic Center black hole Sgr A$^*$, plausibly an
analogue of the (hypothetical) black hole powering the persistent source
near FRB 121102: The Galactic Center pulsar PSR J1745$-$2900 has a rotation
measure similar, and similarly varying, to that of FRB 121102, {
suggesting that this may be a general characteristic of the regions around
massive black holes.}
\subsection{Rarity}
\citet{B19} have shown that the space density of repeating FRB with
characteristics similar to those of FRB121102 is $< 10^{-5}$ Mpc$^{-3}$.
This should be compared to the density of galaxies, $\sim 0.003$ Mpc$^{-3}$
for $L^*$ galaxies and $\sim 0.1$ Mpc$^{-3}$ for all galaxies \citep{C16}.
It implies that the sources of repeating FRB are extraordinarily rare in the
Universe, rarer even than the massive black holes of galactic nuclei.  If
repeating FRB are associated with these black holes another restrictive
condition must also be satisfied.  A plausible candidate is collimation
consistent with the discussion of particle acceleration in magnetic
reconnection in Sec.~\ref{reconnection})---FRB are only observed if the
radiation and the accelerating electric field is pointing towards the Earth
\citep{K17a}.
\subsection{Propagation of radiation in a jet}
{ At a distance $r$ from the source of a jet with power $L_{jet}$ in
particle kinetic energy, the lepton density}
\begin{equation}
	\label{ne}
	n_e = {L_{jet} \over \Omega r^2 \mu \gamma_{lep} c^3},
\end{equation}
where $\Omega$ is the jet's solid angle, $\mu$ is the mass per lepton
(for a purely leptonic jet $\mu = m_e$ but for a baryonic jet $\mu \approx
m_p$) and $\gamma_{lep}$ is the Lorentz factor { of the leptons,
including both bulk jet motion and their disordered motion within the jet}.
For $L_{jet} = 10^{40}\,$ergs/s, $\Omega = 0.01$ and $r = 10 r_{Sch}$, where
$r_{Sch}$ is the Schwarzschild radius and $M = 10^7 M_\odot$, $n_e = 2
\times 10^7/\gamma_{lep}\,$cm$^{-3}$ for a baryonic jet and $n_e = 5 \times
10^{10}\,$cm$^{-3}$ for a leptonic jet.  $L_{jet}$ may be only a small
fraction of the accretion luminosity if much of the power is carried by
Poynting flux rather than by particle kinetic energy.

The plasma frequency in its frame of a jet of relativistic leptons is
\begin{equation}
	\label{nup}
	\nu_p = \sqrt{n_e e^2 \over \pi m_e \gamma_e},
\end{equation}
where $\gamma_e < \gamma_{lep}$ is a mean lepton Lorentz factor in the frame
of the jet.  The values of the parameters are very uncertain but $\gamma_e
\gg 1$ is likely.  The frequency in the observer's frame is blue-shifted
from that in the jet frame (more than outweighing the transformation of
$n_e$, which enters to the $1/2$ power), the observed $\sim 1\,$ GHz
radiation may propagate through such a jet.  This also explains the failure
to detect FRB 121102 at 150 MHz \citep{HSV19} without attributing it to
scattering.

The dependence of $n_e$ (Eq.~\ref{ne}) and hence of $\nu_p$ (Eq.~\ref{nup})
on $L_{jet}$ may explain the low luminosity of the persistent source
associated with the repeating FRB 121102: FRB produced in denser and more
energetic jets cannot emerge to be observed.
\section{Non-Repeating FRB}
\label{nonrep}
The failure to observe any repetition of { apparently non-repeating FRB}
during $4.5 \times 10^7$ s of cumulative observation (\citet{S18}; if all
are drawn from the same ensemble this is equivalent to the same total
duration of observation of a single FRB) suggests that they truly do not
repeat, emitting only one burst in their lifetimes.  Their sources may come
to a catastrophic end, during which the FRB is emitted.  This may be
contrasted with the multiple repetitions of FRB 121102 during some (but not
all) observing runs, { typically five hours long}.

The ASKAP FRB have a mean DM about half that of the Parkes FRB and
indistinguishable from that of the CHIME FRB \citep{LYB19}, although the
ASKAP fluence detection threshold is many times (estimates are 20--50 times)
those of Parkes and CHIME.  In a Euclidean model the mean distance should be
proportional to the $-1/2$ power of the threshold, in disagreement with this
prediction.  Not only are the mean DM from surveys with very different
thresholds so similar, but within the ASKAP, Parkes and UTMOST surveys DM do
not trend systematically with fluence \citep{S18}, which is equivalent to
dividing each survey into sub-populations with different detection
thresholds.

\citet{LYB19} suggest that DM is not proportional to distance, but is
mostly attributable to near-source plasma that is statistically independent
of distance.  This is likely inconsistent with pulsar and SGR models of FRB,
because Galactic pulsars and SGR do not have significant near-source
contributions to DM.  This objection may not apply to pulsars in very young
supernova remnants (SNR), but there are statistical arguments against
attributing FRB DM to SNR \citep{K16a}.  If the suggestion of \citet{LYB19}
is accepted, they infer a bound $z \lesssim 0.02$ on the redshifts of
non-repeating FRB from their bounds on the intergalactic contribution to DM.

An alternative explanation is that the DM are mostly intergalactic and that
the source density or detectability cut off rapidly at $z \gtrsim 1$ as a
result of cosmic evolution and redshift.  This hypothesis cannot be
evaluated quantitatively because we are ignorant of the source population
and its evolution and of the effect of redshift on detectability (the
K-correction of extragalactic optical astronomy), but a simple model is
possible.  We assume the Euclidean inverse square law and geometry, but
impose an abrupt cutoff on detectability at a distance $d_{max}$.

For a distribution $n(L)$ of FRB strength $L$ (which could be energy, 
luminosity, or luminosity times some power of duration---any quantity that
follows an inverse square law, but whose specific definition depends on the
characteristics of the receiving system), we assume $n(L) \propto
L^{-\beta}$ and a detection threshold $F_0$, where $F=L/(4\pi d^2)$.  The
mean distance
\begin{equation}
	\label{dmean}
	\begin{split}
	\langle d \rangle &= {\int_{0}^{d_{max}}\!4 \pi d^3\,dd
	\int_{L=4\pi d^2 F_0}^\infty\!L^{-\beta}\,dL \over
	\int_{0}^{d_{max}}\!4 \pi d^2\,dd
	\int_{L=4\pi d^2 F_0}^\infty\!L^{-\beta}\,dL}\\
	&= d_{max}{5 - 2\beta \over 6 -2\beta},
	\end{split}
\end{equation}
where we have assumed $\beta < 5/2$.  If this condition is not met the
normalizing denominator diverges unless a near-observer cutoff is
introduced, and if $\beta \ge 3$ the numerator also requires such a cutoff.
If $\beta \le 1$ the $L$ integrals diverge, so this result is valid only in
the plausible range $1 < \beta < 5/2$; \citet{LP19} suggest $\beta \approx
1.7$ (if $\beta < 2$ the total FRB energy or luminosity diverge at the high
energy limit of the integral over $L$ and another cutoff must be introduced
in evaluating it).  This result differs from that of \citet{LYB19} because
of the introduction of a cosmological cutoff, violating their Euclidean
assumption.

Eq.~\ref{dmean} explains the near-independence of the mean DM on the
sensitivity of the survey, which now supports the attribution of most of
the dispersion to the intergalactic medium and cosmological distances of
FRB.  If, as must be the case, the cosmological cutoff is not abrupt,
$\langle d \rangle$ will have some dependence on $F_0$, as observed
\citep{S18,LYB19}.

If, instead, there is a characteristic strength $L_0$ with $n(L) \propto
\delta(L-L_0)$ and $L_0 > 4 \pi d_{max}^2 F_0$, then $\langle d \rangle =
3d_{max}/4$.  $\langle d \rangle$ will be the same for all observing systems
that satisfy the inequality.  If $L_0 < 4 \pi d_{max}^2 F_0$ then the $-1/2$
power dependence of $\langle d \rangle$ on $F_0$ \citep{LYB19} is recovered.

The empirical near-independence of $DM$ (approximately proportional to $d$
for $z \lesssim 1$) of $F_0$, is consistent with either a power law
distribution of $L$ or a distribution peaked at a strength sufficient for
observation out to $z \sim 1$.  The fact that the ASKAP detection rate of
$1.4 \times 10^4$/sky-y \citep{S18} is nearly two orders of magnitude less
than the rate (in the same frequency band) of the more sensitive Parkes
survey of $\sim 10^6$/sky-y \citep{L18} points toward the power law, or some
other broad, distribution.  However, rejection of the $\delta$-function
distribution does not contradict the inference (Sec.~\ref{sky}) of a
characteristic scale of strength of non-repeating FRB.  
\subsection{Collapsing Neutron Stars}
\label{nstobh}
Non-repeating FRB may involve the destruction of their progenitors.
\citet{FR14} suggested that FRB are produced by the collapse of rotationally
stabilized neutron stars with masses greater than the maximum mass of a
non-rotating neutron star after they lose their angular momentum.  Yet there
is no { compelling} evidence that rapidly rotating strongly magnetized
neutron stars exist or can be made, and Appendix~\ref{origspin} { argues
against their formation}.

Known rapidly rotating neutron stars (millisecond pulsars) have small
magnetic moments $\mu \lesssim 10^{27}\,$gauss-cm$^3$, and magnetic energy
$\lesssim 10^{36}$ ergs, insufficient to power FRB even with unit
efficiency.  Their rapid rotation is attributed to ``recycling'' by
accretion.  In addition, a very rapidly rotating { strongly magnetized}
neutron star drives away surrounding matter { with its relativistic
wind}, so there may be no nearby particles to accelerate.  Extrapolating its
spindown backward in time indicates that the Crab pulsar was born with a
period of about 17 ms, too slow for rotation to be significantly
stabilizing, and there is no evidence that other neutron stars are born
faster.

I therefore consider { neutron star} collapse produced by accretion.
Accretion from binary companions powers many Galactic X-ray sources,
permitting an estimate of the rate at which neutron stars will be pushed
over their stability limit.  A neutron star born with a mass of
$1.2 M_\odot$ must accrete about $1 M_\odot$ to collapse \citep{A13},
consistent with the life expectancy and accretion rates of at least some
X-ray binaries \citep{TvdH06}.  The more massive ($\approx 2 M_\odot$)
observed neutron stars may have been formed with those masses (requiring
less accretion to collapse) or be the result of accretion so far
insufficient for collapse.
\subsubsection{Rate}
\label{rate}
An accreting neutron star in a binary system may accumulate mass until it
exceeds its maximum mass, when it will rapidly collapse to a black hole.
Such neutron stars emit their accretional energy as X-rays, so that the
total rate of neutron star accretion in a galaxy is bounded by its X-ray
luminosity (there are other contributors to X-ray luminosity, including
supernova remnants and accreting black holes).  If a mass $\Delta M$ must be
accreted to push a neutron star over its maximum mass and the accretion
energy per unit mass is $\epsilon c^2$ then the mean repetition time between
collapses { in a galaxy} is
\begin{equation}
	t_{rep} = {\Delta M \epsilon c^2 \over L_{XNS}},
\end{equation}
where $L_{XNS}$ is the luminosity attributable to accretion onto neutron
stars.  For a nominal { galactic} $L_{XNS} = 10^{41}\,$ ergs/s
\citep{KF04}, $\Delta M = 1 M_\odot$ and $\epsilon c^2 = 10^{20}\,$ergs/g,
$t_{rep} \sim 10^5\,$y.  This value may be barely consistent with the
observed FRB rate of $\sim 10^6$/sky-y.  $L_{XNS}$ may be much greater in
star-forming galaxies \citep{MGS12}.  For our Galaxy $L_{XNS} \approx 3
\times 10^{39}\,$ ergs/s \citep{KF04} and $t_{rep} \sim 3 \times 10^6\,$y.

This explains why collapse FRB are rare, and why their observed distribution
is not dominated by a Galactic component.  In this model FRB radiation
cannot be collimated, for that would require a collapse rate much higher
than the observed FRB rate, { but the enormous energy available in
collapse obviates the need for collimation}.  If an average could be taken
over a time $t_{av} \gg t_{rep}$ the Galactic plane would dominate the FRB
sky { just as it dominates the visible/IR sky (Sec~\ref{sky}), because in
this model non-repeating FRB are products of stellar evolution}.
\subsubsection{Energetics}
\label{nsenergetics}
%
%
Relativistic MHD calculations \citep{L12,D13} of collapsing magnetized
neutron stars { in vacuum} have indicated radiation at frequencies $\sim
10^4\,$Hz of 5--16\% of the initial magnetostatic energy.  The external
magnetostatic energy (ignoring relativistic effects) is $\mu^2/3r^3 \approx
3 \times 10^{41} \mu_{30}^2\,$ergs).  To explain non-repeating FRB with
estimated energies $\sim 10^{40}$ ergs by this process requires at least one
of: 1) collimation (excluded by the argument of Sec.~\ref{rate}); 2) fairly
efficient conversion of the collapse radiation to FRB; or 3) $\mu_{30} \gg
1$.

Accreting neutron stars in binary systems, observed as X-ray sources,
typically have magnetic moments $\mu = 10^{30}\text{--}10^{31}
\,$gauss-cm$^3$.  The magnetic moments of SGR are believed
\citep{K82,TD92,TD95} to be in the range $\mu = 10^{32}\text{--}10^{33}
\,$gauss-cm$^3$.  Although no known SGR is in a binary system, this
establishes the existence of neutron stars with extremely large magnetic
moments and magnetostatic energies, some of which may have, { perhaps
later in their evolution}, mass-transferring binary companions.
\subsection{Collapsing White Dwarfs}
\label{wdtons}
{ Accretion-induced collapse (AIC) of white dwarfs to neutron stars has
long been discussed as a possible origin of neutron stars and as a possible
mechanism of supernov\ae.  Many single magnetic white dwarfs are unusually
massive, some with masses $> 1.3 M_\odot$ \citep{FdMG15} that may be the
result of dissolution of binary companions as their mass is transferred.
``Polars'' (synchronously rotating magnetic white dwarfs in mass transfer
binaries) have been observed with fields up to 200 MG.  Accretion-induced
collapse of strongly magnetized white dwarfs is plausible, although no
system is known in which such collapse can be predicted.
\subsubsection{Rate}
Accretion onto white dwarfs is $\sim \text{100--1000}$ times less luminous
than accretion onto neutron stars, the value of $\sim 100$ only applying at
a radius of $\sim 10^8\,$cm, near the threshold of collapse.  As a result,
measurements of the X-ray luminosity of galaxies do not constrain the rate
of accretion-induced collapses of white dwarfs.  Their low luminosity also
makes it hard to identify candidate objects in our Galaxy.
\subsubsection{Energetics}
\label{wdenergetics}
During pre-collapse contraction (a slow process resulting from the increase
in mass) as well as dynamical collapse itself a frozen-in magnetic field
increases $B \propto r^{-2}$, so the magnetic moment decreases $\mu \sim
\mu_0 r/r_0$, where $\mu_0$ and $r_0$ are the initial moment and radius.
This changing dipole moment radiates during dynamic collapse, principally
at its end, as the star approaches neutron star density.  At a radius $r$,
time scale $\Delta t \sim r/v$, where $v$ is the infall speed, and frequency
$\omega \sim 1/\Delta t$, the radiated energy
\begin{equation}
	{\cal E} \sim {\omega^4 \mu^2 \over 3c^3} \Delta t \sim {\mu_0^2 r^2
	\over 3 r_0^2 (c \Delta t)^3} \sim {\mu_0^2 \over r_0^2} \left({v
	\over c}\right)^2 {1 \over 3 c \Delta t}.
\end{equation}
If the field is frozen-in during the pre-collapse quasi-static evolution of
the white dwarf, $\mu_0$ and $r_0$ may have the values observed in magnetic
white dwarfs of $\mu_0 \sim 10^{35}$ gauss-cm$^3$ and $r_0$ as small as $5
\times 10^8$ cm.  Rare, as yet unobserved, objects may have larger ratios
$\mu_0/r_0$.

Non-repeating FRB with width $\Delta t \sim 1\,$ms must be produced in the
last ms (or less) of the collapse.  \citet{F18,CHIME1} reported
non-repeating FRB with $\Delta t \sim 0.1\,$ms, a brevity previously only
found in repeating FRB \citep{Mi18}, although it is not impossible that
these ultra-short events are repeating FRB whose repetitions have not been
observed.  In the final stages of collapse $v \sim 0.5c$, so that
\begin{equation}
	\label{energy}
	{\cal E} \sim 10^{44} \left({\mu_0/r_0 \over 2 \times 10^{26}
	\text{gauss-cm}^2}\right)^2 {1\,\text{ms} \over \Delta t}\
	\text{ergs},
\end{equation}
providing enough low frequency (kHz) energy to power the most energetic
($\sim 10^{40}$ ergs) FRB even with a low ($\sim 10^{-4}$) efficiency of
conversion to GHz radiation.
\subsection{Time Scale}
For either collapse model to explain non-repeating FRB, not only must the
low frequency (kHz) radiation produced by the changing dipole moment of the
collapsing object be emitted within the brief duration of the FRB, but so
must the GHz radiation defining the FRB itself.  This might seem to require
an emission region of size $< c \Delta t \sim 3 \times 10^6\text{--}3 \times
10^7\,$cm, corresponding to $\Delta t \sim 0.1\text{--}1\,$ms.  This would
be an objection to this hypothesis because it would require the presence of
radiating matter with a region that had (as part of a collapsing white
dwarf) presumptively collapsed to the neutron star, or (as part of a
collapsing neutron star) had been inside a plausible accretion disc.

Instead, I suggest that when the energetic (Eq.~\ref{energy}) pulse of
dipole radiation encounters surrounding matter particles are accelerated in
the direction of the Poynting vector, radially outward.  At a radius $R$
particles of Lorentz factor $\gamma$ effectively radiate to the observer
only from a patch of width $\sim R/\gamma$, and the resulting $\Delta t \sim
R/(c\gamma^2)$.  This cannot be evaluated quantitatively without an estimate
of $\gamma$, but is consistent with $R$ comparable to the size of mass
transfer binaries ($R \sim 10^{11}\,$cm) if $\gamma \sim 10^2$.}
\section{Conclusions}
\label{conclusions}
Two models are required, one for repeating FRB and another for non-repeating
FRB, that appear to be different phenomena.  { This is inferred from
several arguments, the most compelling of which are duty factors and
rotation measures \citep{Mi18}, each of which differ by at least three
orders of magnitude between the repeaters and apparent non-repeaters.  The
second repeater \citep{CHIME2} may soon provide a test of this inference if
its duty factor and rotation measure are measured.  The two models are
distinct, one involving accreting black holes to describe repeating FRB and
the other, that can explain only non-repeating FRB, involving neutron star
or white dwarf collapse.}

In the { model of repeating} FRB, they are produced by the highly
collimated emission of very relativistic electrons accelerated by magnetic
reconnection, as discussed in Sec.~\ref{reconnection}.  This may occur in
shear flows in jets accelerated by accretion onto massive black holes.
Accretion discs and their jets are long-lived and fluctuate, consistent with
the { long life and episodic behavior of the repeating} FRB 121102.  This
model builds on elements of earlier speculations that FRB are wandering
narrow beams \citep{K17a} and that they are associated with intermediate
mass black holes \citep{C17,M17,K17c}.  Because the electron density in
accretion discs is high, GHz radiation cannot propagate through them, and
must be produced in and escape through the low density relativistic jet.

The distribution of non-repeating FRB on the sky implies either that they
are not associated with a stellar population or that they are rare,
energetic and singular events, {\it none\/} of which have occurred in the
Galaxy in the regions and during the time in which they could have been
observed.  This distinguishes them from phenomena, like the giant pulses of
pulsars or the outbursts of the repeating FRB 121102, that have a broad
distribution of strengths and whose energy output is dominated by the
weakest events.

The absence of repetitions during the monitoring of ``non-repeating'' FRB
sets an upper bound on their mean repetition rate.  If repetitions are
Poissonian this upper bound is a few times the reciprocal of the {\it
cumulative\/} { } monitoring time.  The ASKAP bound \citep{S18} should be
interpreted with caution because of its high detection threshold of about 40
Jy-ms; if these sources do repeat, the repetition rate would be higher for
lower thresholds unless they are standard flashbulbs.  It also cannot be
directly compared to the characteristic interval time of FRB 121102 of 60 s
(Fig.~\ref{intervals}) because of the lower detection threshold (about 20
Jy-$\mu$s) and different frequency (4--8 GHz) of \citet{Z18}.  If
repetitions are positively correlated (as they are in FRB 121102) the upper
bound on the {\it mean\/} repetition rate is even less, its value depending
on the shape of the correlation function, but if they are anti-correlated it
could be relaxed.

Collapsing { white dwarfs and} neutron stars are distributed throughout
galaxies as were their progenitors, Population I stars.  As a result, a
large RM like that of PSR J1745$-$2900 in the region around a galactic
nucleus like Sgr A$^*$ is not expected, consistent with the comparatively
low RM of non-repeating FRB \citep{Mi18}.  This spatial distribution does
not contradict the absence of FRB in the Galactic plane (Sec.~\ref{sky})
because { non-repeating} FRB have a large characteristic strength and are
very infrequent.  In a sufficiently long time average the Galactic plane
would dominate the { non-repeating} FRB sky (because of the proximity of
Galactic FRB progenitors), but { ``sufficiently long'' is longer than the
interval between non-repeating FRB in a galaxy, $\sim 10^5\,$y}.  FRB
described by this model do not repeat, will generally not be found in
galactic nuclei, and are likely to be identified with galaxies and regions
with rapid star formation.  Their accompanying gravitational wave signal is
weak because slowly rotating stars collapse almost spherically
symmetrically.

{ The hypotheses and models presented here that repeating and apparently
non-repeating FRB are fundamentally different phenomena predict that
repeaters discovered in the future will qualitatively resemble FRB 121102 in
several respects:  Their bursts will be (compared to most non-repeaters) of
low energy, they will have high but fluctuating rotation measure, if in
galaxies of regular form they will be in the host's central region, they may
be associated with weak (but not strong) persistent radio sources, some or
all of their bursts will show downward spectral drifts, and their
repetitions will not be periodic.  It also predicts that extended
observation will divide bursters into two distinct classes on the basis of
their duty factor; apparent non-repeaters will either remain non-repeaters
indefinitely or will repeat comparatively frequently (of course, every
repeater is first discovered as a single burst).}

\appendix
\section{Original Spin}
\label{origspin}
{ It is possible to estimate the maximum angular spin rate $\omega_0$ of
a neutron star newly formed by a core collapse supernova by equating the
accretional torque with the spindown torque of a magnetic rotor immersed in
a medium with Alfv\'en speed $v_A$:
\begin{equation}
	{\dot M}\ell \sim {\omega^3 \mu^2 \over v_A^3},
\end{equation}
where $\ell = 10^{16}\,$cm$^2$/s is the specific angular momentum of
accreting matter and $v_A \sim B/\sqrt{\rho} \sim \mu/\sqrt{Mr^3}$, where
during the formation of the neutron star the density $\rho \sim M/r^3$.
This yields
\begin{equation}
	\omega_0 \sim \left({{\dot M} \ell \mu \over M^{3/2} r^{9/2}}
	\right)^{1/3}.
\end{equation}
The accretion rate $\dot M \sim M/t_{acc}$, where the accretion time is
poorly known.  The rate of rise of the SN1987A neutrino flux indicates that
$t_{acc} < 0.2\,$s, but this estimate is limited by the low count rate and
$t_{acc}$ could be much less than the indicated upper bound.  Numerically,
taking $r = 10^6\,$cm, the final neutron star radius, to maximize
$\omega_0$,
\begin{equation}
	\omega_0 \sim \left({\mu_{30} \over t_{acc\,-3}}\right)^{1/3}\ 
	60\,\text{s}^{-1},
\end{equation}
where $\mu_{30} \equiv \mu/10^{30}\,$gauss-cm$^3$ and $t_{acc\,-3} \equiv
t_{acc}/10^{-3}\,$s.  This rate is several times less than the inferred
original spin of the Crab pulsar $\omega_0 \approx 400\,$/s, but is
consistent with the longer initial periods indicated by statistical studies
of pulsars \citep{N13}.  Fallback may further spin up the star, but this
crude estimate is an argument against the suggestion of neutron stars born
spinning near break-up, as required in pulsar models of repeating FRB
\citep{K18a}.}
\section{Duty Factor}
\label{duty}
{ Assume a power law distribution of spectral densities $\cal S$
\begin{equation}
	\label{powerlaw}
	{dN \over d{\cal S}} = f {\cal S}^{-\beta}.
\end{equation}
The number of temporal intervals in an observation of duration $T$ and
temporal resolution $\Delta t$ with flux density ${\cal S}$ greater than a
threshold ${\cal S}_{min}$ (corresponding to the number of bursts if they
are not temporally resolved, or if the resolution is artifically broadened
to avoid resolving them) is, if $\beta > 1$,
\begin{equation}
	\label{N}
	N = {T \over \Delta t} \int_{{\cal S}_{min}}^\infty\!{dN \over
	d{\cal S}}\,d{\cal S} =
	{T \over \delta t}{f \over \beta - 1} {\cal S}_{min}^{1 - \beta}.
\end{equation}
Setting $N = N_b$, the number of observed bursts
\begin{equation}
	f = N_b {\Delta t \over T}
	{\beta - 1 \over {\cal S}_{min}^{1 - \beta}}.
\end{equation}
The expected brightness of the brightest burst may be estimated by taking
$N = 1$ in Eq.~\ref{N}:
\begin{equation}
	\label{Smax}
	{\cal S}_{max} \sim {\cal S}_{min} N_b^{1/(\beta - 1)}
\end{equation}
and the typical brightness ${\cal S}_0 \ll {\cal S}_{min}$ in intervals in
which no bursts are detected by taking $N = T/\Delta t$
\begin{equation}
	{\cal S}_0 \sim {\sim S}_{min}
	\left({N_b \over T/\Delta t}\right)^{1/(\beta - 1)}.
\end{equation}
These extrapolations are only as good as the assumption of the power law
distribution Eq.~\ref{powerlaw}.

The means in Eq.~\ref{dutyfactor} are taken over the time of observation
(not over bursts), so that for $2 > \beta > 1$
\begin{equation}
	\begin{split}
	\langle {\cal S} \rangle &= \int_{{\cal S}_{min}}^{{\cal S}_{max}}\!
	{dN \over d{\cal S}}{\cal S}\,d{\cal S}\\ &= N_b {\Delta t \over T}
	{\beta - 1 \over 2 - \beta} {{\cal S}_{max}^{(2-\beta)} \over
	{\cal S}_{min}^{(1-\beta)}} = {\Delta t \over T}
	{\beta - 1 \over 2 - \beta} {\cal S}_{min} N_b^{1/(\beta-1)}
	\end{split}
\end{equation}
and
\begin{equation}
	\begin{split}
	\langle {\cal S}^2 \rangle &= \int_{{\cal S}_{min}}^{{\cal S}_{max}}\!
	{dN \over d{\cal S}}{\cal S}^2\,d{\cal S}\\ &= N_b {\Delta t \over T}
	{\beta - 1 \over 3 - \beta} {{\cal S}_{max}^{(3-\beta)} \over
	{\cal S}_{min}^{(1-\beta)}} = {\Delta t \over T}
	{\beta - 1 \over 3 - \beta} {\cal S}_{min}^2 N_b^{2/(\beta-1)}.
	\end{split}
\end{equation}
Then
\begin{equation}
	\label{dutypowerlaw}
	D = {\langle{\cal S}\rangle^2 \over \langle{\cal S}^2\rangle} =
	{\Delta t \over T}{(\beta - 1)(3 - \beta) \over (2 - \beta)^2}
	\approx 10 {\Delta t \over T},
\end{equation}
for $\beta \approx 1.7$ \citep{G19,LP19}.
These results assume that no bursts are observed with ${\cal S}$ much
greater than the nominal ${\cal S}_{max}$ (Eq.~\ref{Smax}); the means
$\langle {\cal S} \rangle$ and $\langle {\cal S}^2 \rangle$ are not properly
defined for $\beta < 2$ and $\beta < 3$, respectively, because if their
upper bounds are taken as infinity the integrals are dominated by rare
strong outliers and diverge.  However, most data realizations have no such
outliers, justifying the imposition of the upper bound ${\cal S}_{max}$ when
considering a single finite data set.

This result is independent of the detection threshold ${\cal S}_{min}$, so
that results from different telescopes with different values of
${\cal S}_{min}$ can be compared (provided the power law assumption is
valid), as well as results obtained at different frequencies.  If more than
one burst is observed then $D$ can be estimated, at least roughly, from the
data.  If only one burst is observed only an upper bound to $D$ exists and
Eq.~\ref{dutypowerlaw} can be used to interpret this as a rough lower bound
on the repetition time $T$.}

\bsp 
\label{lastpage} 

\begin{thebibliography}{99}
\bibitem[\protect\citeauthoryear{Antoniadis {\it et al.\/}}{2013}] {A13}
	Antoniadis, J., Freire, P. C. C., Wex, N. {\it et al.\/} 2013
		Science 340, 448.
\bibitem[\protect\citeauthoryear{Bassa {\it et al.\/}}{2017}]{B17} Bassa, C.
G., Tendulkar, S. P., Adams, E. A. K. {\it et al.\/} 2017 \apj\ 843, L8.
\bibitem[\protect\citeauthoryear{Bhandari {\it et al.\/}}{2018}]{B18}
	Bhandari, S., Keane, E.~F., Barr, E.~D. {\it et al.\/} 2018 \mnras\
	475, 1427.
\bibitem[\protect\citeauthoryear{Bhandari {\it et al.\/}}{2019}]{B19}
	Bhandari, S., Bannister, K. W., James, C. W. {\it et al.\/} 2019
		arXiv:1903.06525.
\bibitem[\protect\citeauthoryear{Bower {\it et al.\/}}{2015}]{B15} Bower,
	G.~C., Deller, A., Demorest, P. {\it et al.\/} 2015 \apj\ 798, 120.
\bibitem[\protect\citeauthoryear{Bower {\it et al.\/}}{2018}]{BBD18} Bower,
	G.~C., Broderick, A., Dexter, J. {\it et al.\/} 2018 \apj\ 868, 101.
\bibitem[\protect\citeauthoryear{Caleb, Spitler \& Stappers}{2018}]{CSS18}
	Caleb, M., Spitler, L. G. \& Stappers, B. W. 2018 Nature Astronomy
	2, 839
\bibitem[\protect\citeauthoryear{Chatterjee {\it et al.\/}}{2017}]{C17}
Chatterjee, S., Law, C. J., Wharton, R. S. {\it et al.\/} 2017 Nature 541,
58.
\bibitem[\protect\citeauthoryear{CHIME/FRB}{2019a}]{CHIME1} CHIME/FRB
	Collaboration 2019a Nature doi:10.1038/s41586-018-0864-x
		arXiv:1901.04525.
\bibitem[\protect\citeauthoryear{CHIME/FRB}{2019b}]{CHIME2} CHIME/FRB
	Collaboration 2019b Nature doi:10.1038/s41586-018-0867-7
		arXiv:1901.04524.
\bibitem[\protect\citeauthoryear{Conselice {\it et al.\/}}{2016}]{C16}
	Conselice, C. J., Wilkinson, A., Duncan, K. \& Mortlock, A. 2016
		\apj\ 830, 83.
\bibitem[\protect\citeauthoryear{Cordes {\it et al.\/}}{2004}]{CBH04}
	Cordes, J. M., Bhat, N. D. R., Hankins, T. H. {\it et al.\/} 2004
	\apj\ 612, 375.
\bibitem[\protect\citeauthoryear{Cui, B. \& McLauglin, M.}{2018}]{CM16}
	Cui, B. \& McLaughlin, M. \url{http://astro.phys.wvu.edu/rratolog/}
	accessed Sept. 6, 2018.
\bibitem[\protect\citeauthoryear{Desvignes {\it et al.\/}}{2018}]{D18}
	Desvignes, G., Eatough, R.~P., Pen, U.~L. {\it et al.\/} 2018 \apj\
	852, L12.
\bibitem[\protect\citeauthoryear{Dionysopoulou {\it et al.\/}}{2013}]{D13}
	Dionysopoulou, K., Alec, D., Palenzuela, C., Rezzolla, L. \&
	Giaccomazzo, B. 2013 \prd\ 88, 4020.
\bibitem[\protect\citeauthoryear{Dom\'{i}nguez {\it et al.\/}}{2011}]{D11}
Dom\'{i}nguez, A., Primack, J. R., Rosario, D. J. {\it et al.\/} 2011
\mnras\ 410, 2256.
\bibitem[\protect\citeauthoryear{Falcke \& Rezzolla}{2014}]{FR14} Falcke,
	H. \& Rezzolla, L. 2014 \aap\ 562A, 137.
\bibitem[\protect\citeauthoryear{Farah, Flynn, Bailes {\it et al.\/}}{2018}]
	{F18} Farah, W., Flynn, C., Bailes, M. {\it et al.\/} 2018 \mnras\
		478, 1209.
\bibitem[\protect\citeauthoryear{Ferrario, de Martino \& G\"{a}nsicke}
	{2015}] {FdMG15} Ferrario, L, de Martino, D. \& G\"{a}nsicke, B.
		2015 Sp.~Sci.~Rev. 191, 111.
\bibitem[\protect\citeauthoryear{Gajjar {\it et al.}}{2018}]{G18} Gajjar,
V., Siemion, A. P. V., Price, D.~C. {\it et al.\/} 2018 \apj\ 863, 2.
\bibitem[\protect\citeauthoryear{Gonzalez \& Parker}{2016}]{GZ16} Gonzalez,
	W. \& Parker, E. eds. 2016 Magnetic Reconnection: Concepts and
		Applications (Springer, New York).
\bibitem[\protect\citeauthoryear{Gourdji {\it et al.\/}}{2019}]{G19}
	Gourdji, K., Michilli, D., Spitler, L. G. {\it et al.\/} 2019
		arXiv:1903.02249.
\bibitem[\protect\citeauthoryear{Hankins \& Eilek}{2007}]{HE07}
	Hankins, T.~H. \& Eilek, J.~A. 2007 \apj\ 670, 693.
\bibitem[\protect\citeauthoryear{Hardy {\it et al.\/}}{2017}]{H17} Hardy,
L. K., Dhillon, V. S., Spitler, L. G. {\it et al.\/} 2017 \mnras\ 472, 2800
\bibitem[\protect\citeauthoryear{Hessels {\it et al.\/}}{2019}]{H19}
	Hessels, J. W. T., Spitler, L. G., Seymour, A. D. {\it et al.\/}
		2019 \apj\ submitted arXiv:1811.10748.
\bibitem[\protect\citeauthoryear{Houben {\it et al.}}{2019}]{HSV19} Houben,
	L.~J.~M., Spitler, L.~G., ter Veen, S. {\it et al.\/} 2019
		arXiv:1902.01779.
\bibitem[\protect\citeauthoryear{James {\it et al.\/}}{2019}]{J18} James, C.
W., Ekers, R. D., Macquart, J.-P., Bannister, K. W. \& Shannon, R. M. 2019
\mnras\ 483, 1342 arXiv:1810.04357.
\bibitem[\protect\citeauthoryear{Katz}{1982}]{K82} Katz, J. I. 1982
	\apj\ 260, 371.
\bibitem[\protect\citeauthoryear{Katz}{2002}]{K02} Katz, J. I. 2002
	{\it The Biggest Bangs\/} Oxford U.~Press, New York.
\bibitem[\protect\citeauthoryear{Katz}{2014}]{K14} Katz, J. I. 2014
	\prd\ 89, 103009.
\bibitem[\protect\citeauthoryear{Katz}{2016a}]{K16a} Katz, J. I. 2016a
	\apj\ 818, 19.
\bibitem[\protect\citeauthoryear{Katz}{2016b}]{K16b} Katz, J. I. 2016b
        \apj\ 826, 226.
\bibitem[\protect\citeauthoryear{Katz}{2017a}]{K17a} Katz, J. I. 2017a
	\mnras\ 467, L96.
\bibitem[\protect\citeauthoryear{Katz}{2017b}]{K17b} Katz, J. I. 2017b
	\mnras\ 469, L39.
\bibitem[\protect\citeauthoryear{Katz}{2017c}]{K17c} Katz, J. I. 2017c
	\mnras\ 471, L92.
\bibitem[\protect\citeauthoryear{Katz}{2018a}]{K18a} Katz, J. I. 2018a Prog.
	Part. Nucl. Phys. 103, 1 doi:10.1016/j.ppnp.2018.07.001
	arXiv:1804.09092. 
\bibitem[\protect\citeauthoryear{Katz}{2018b}]{K18b} Katz, J. I. 2018b 
	\mnras\ 481, 2946 arXiv:1803.01938. 
\bibitem[\protect\citeauthoryear{Katz}{2018c}]{K18c} Katz, J. I. 2018c
	\mnras\ 476, 1849. 
\bibitem[\protect\citeauthoryear{Kazantsev \& Potapov}{2017}]{KP17}
	Kazantsev, A. N. \& Potapov, V. A. 2017 Astron.~Rep. 61, 747.
\bibitem[\protect\citeauthoryear{Kim \& Fabbiano}{2004}]{KF04} Kim, D.-W.
	\& Fabbiano, G. 2004 \apj\ 611, 864.
\bibitem[\protect\citeauthoryear{Kulkarni}{2018}]{Ku18} Kulkarni, S. 2018
	\nat\ Astron. 2, 832.
\bibitem[\protect\citeauthoryear{Kumar, Lu \& Bhattacharya}{2017}]{KLB17}
	Kumar, P., Lu, W. \& Bhattacharya, M. 2017 \mnras\ 468, 2726.
\bibitem[\protect\citeauthoryear{Law {\it et al.\/}}{2017}]{L17} Law, C. J.
	Abruzzo, M. W., Bassa, C. G. {\it et al.\/} 2017 \apj\ 850, 76.
\bibitem[\protect\citeauthoryear{Lehner {\it et al.\/}}{2012}]{L12} Lehner,
	L., Palenzuela, C., Liebling, S. L., Thompson, C. \& Hanna, C. 2012
	\prd\ 86, 104035.
\bibitem[\protect\citeauthoryear{Leinert {\it et al.\/}}{1998}]{L98}
Leinert, Ch., Bowyer, S., Haikala, L. K. {\it et al.\/} 1998 \aaps\ 127,
1.
\bibitem[\protect\citeauthoryear{Lewis, Antiochos \& Drake}{2011}]{LAD11}
	Lewis, W., Antiochos, S. \& Drake, J. eds. 2011 Magnetic
	Reconnection: Theoretical and Observational Perspectives (Springer,
	New York).
\bibitem[\protect\citeauthoryear{Li, Yalinewich \& Breysse}{2019}]{LYB19}
	Li, D. Z., Yalinewich, A. \& Breysse, P. C. 2019 arXiv:1902.10120.
\bibitem[\protect\citeauthoryear{Li {\it et al.\/}}{2011}]{L11} Li, W.,
	Chornock, R, Leamon, J. {\it et al.\/} 2011 \mnras\ 412, 1473.
\bibitem[\protect\citeauthoryear{Lorimer}{2018}]{L18} Lorimer, D. R. 2018
	Nature astronomy 2, 860.
\bibitem[\protect\citeauthoryear{Lu \& Piro}{2019}]{LP19} Lu, W. \& Piro,
	A. L. 2019 arXiv:1903.00014.
\bibitem[\protect\citeauthoryear{Macquart \& Ekers}{2018a}]{ME18a} Macquart,
	J.-P. \& Ekers, R. D. 2018a \mnras\ 474, 1900.
\bibitem[\protect\citeauthoryear{Macquart \& Ekers}{2018b}]{ME18b} Macquart,
	J.-P. \& Ekers, R. D. 2018b \mnras\ 480, 4211.
\bibitem[\protect\citeauthoryear{Marcote {\it et al.}}{2017}]{M17} Marcote,
	B., Paragi, Z., Hessels, J. W. T. {\it et al.} 2017 \apjl\ 834, L8.
\bibitem[\protect\citeauthoryear{McKee {\it et al.\/}}{2018}]{MSB18} McKee,
	J. W., Stappers, B. W., Bassa, C. G. {\it et al.\/} 2018
	arXiv:1811.02856
\bibitem[\protect\citeauthoryear{McLaughlin {\it et al.\/}}{2006}]{M06}
	McLaughlin, M.~A., Lyne, A.~G., Lorimer, D.~R. {\it et al.\/} 2006
	\nat\ 439, 817.
\bibitem[\protect\citeauthoryear{Michilli {\it et al.\/}}{2018}]{Mi18}
	Michilli, D., Seymour, A., Hessels, J. W. T. {\it et al.\/} 2018
	\nat\ 553, 182.
\bibitem[\protect\citeauthoryear{Mineo, Gilfanov \& Sunyaev}{2012}]{MGS12}
	Mineo, S., Gilfanov, M. \& Sunyaev, R. 2012 \mnras\ 419, 2095.
\bibitem[\protect\citeauthoryear{Noutsos {\it et al.\/}}{2013}]{N13}
	Noutsos, A., Schnitzeler, D. H. F. M., Keane, E. F., Kramer, M. \&
	Johnston, S. 2013 \mnras\ 430, 2281.
\bibitem[\protect\citeauthoryear{Palaniswamy, Li \& Zhang}{2018}]{PLZ18}
	Palaniswamy, D., Li, Y. \& Zhang, B. 2018 \apjl\ 854, L12. 
\bibitem[\protect\citeauthoryear{Pasetto {\it et al.\/}}{2016}]{P16}
	Pasetto, A., Kraus, A., Mack, K.-H. {\it et al.\/} 2016 \aap\ 586,
	117.
\bibitem[\protect\citeauthoryear{Petroff {\it et al.\/}}{2015}]{P15}
	Petroff, E., Bailes, M., Barr, E. D. {\it et al.\/} 2015 \mnras\
	447, 246.
\bibitem[\protect\citeauthoryear{Petroff {\it et al.\/}}{2016}]{frbcat}
	Petroff, E., Barr, E. D., Jameson, A. {\it et al.\/} 2016 Pub.
	Astr. Soc. Australia 33, e045 \url{www.frbcat.org} accessed Feb.
	14, 2019.
\bibitem[\protect\citeauthoryear{Petroff {\it et al.\/}}{2017}]{P17}
	Petroff, E., Burke-Spolaor, S., Keane, E.~F. {\it et al.\/} 2017
	\mnras\ 469, 4465.
\bibitem[\protect\citeauthoryear{Philippov {\it et al.\/}}{2019}]
	{PUSC19} Philippov, A., Uzdensky, D. A., Spitkovsky, A. \& Cerutti,
	B. 2019 arXiv:1902.07730.
\bibitem[\protect\citeauthoryear{Platts {\it et al.\/}}{2018}]{P18} Platts,
E., Weltman, A., Walters, A., Tendulkar, S. P., Gordin, E. B. \& Kandhai, S.
2018 Physics Reports in press (arXiv:1810.05836)
\url{http://frbtheorycat.org}.
\bibitem[\protect\citeauthoryear{Popov \& Stappers}{2007}]{PS07} Popov, M.
	V. \& Stappers, B. 2007 \aap\ 470, 1003.
\bibitem[\protect\citeauthoryear{Romero {\it et al.\/}}{2016}]{RdVV16}
	Romero, G. E., del Valle, M. V. \& Vieyro, F. L. 2016 \prd\ 93,
	023001.
\bibitem[\protect\citeauthoryear{Scholz {\it et al.\/}}{2016}]{Sc16} Scholz,
	P., Spitler, L.~G., Hessels, J.~W.~T. {\it et al.\/} 2016 \apj\
	833, 177.
\bibitem[\protect\citeauthoryear{Scholz {\it et al.\/}}{2017}]{S17} Scholz,
P., Bogdanov, S., Hessels, J. W. T. {\it et al.\/} 2017 \apj\ 846, 80.
\bibitem[\protect\citeauthoryear{Shannon {\it et al.\/}}{2018}]{S18}
Shannon, R. M., Macquart, J.-P., Bannister, K. W. {\it et al.\/} 2018
\nat\ 562, 386.
\bibitem[\protect\citeauthoryear{Soglasnov {\it et al.\/}}{2004}]{S04}
	Soglasnov, V.~A., Popov, M.~V., Bartel, N., Cannon, W., Novikov, A.~Y., Kondratiev, V.~I. \& Altunin, V.~I. 2004 \apj\ 616, 439.
\bibitem[\protect\citeauthoryear{Spitler {\it et al.\/}}{2016}]{Sp16}
	Spitler, L.~G., Scholz, P., Hessels, J.~W.~T. {\it et al.\/} 2016
	\nat\ 531, 202
\bibitem[\protect\citeauthoryear{Tauris \& van den Heuvel}{2006}]{TvdH06}
	Tauris, T. M. \& van den Heuvel, E. 2006 Formation and evolution of
	compact stellar X-ray sources, Compact Stellar X-Ray Sources eds.
	W. Lewin \& M. van der Klis (Cambridge Astrophysics Series 39, 623).
\bibitem[\protect\citeauthoryear{Tendulkar {\it et al.\/}}{2016}]{TKP16}
	Tendulkar, S.~P., Kaspi, V.~M. \& Patel, C. 2016 \apj\ 827, 59.
\bibitem[\protect\citeauthoryear{Tendulkar {\it et al.\/}}{2017}]{T17}
Tendulkar, S. P., Bassa, C. G., Cordes, J. M. {\it et al.\/} 2017 \apj\ 834,
L7.
\bibitem[\protect\citeauthoryear{Thompson \& Duncan}{1992}]{TD92} Thompson,
	C. \& Duncan, R. C. 1992 \apjl\ 392, L9.
\bibitem[\protect\citeauthoryear{Thompson \& Duncan}{1995}]{TD95} Thompson,
	C. \& Duncan, R. C. 1995 \mnras\ 275, 255.
\bibitem[\protect\citeauthoryear{Thornton {\it et al.\/}}{2013}]{T13}
Thornton, D, Stappers, B., Bailes, M. {\it et al.\/} 2013 \sci\ 341, 53.
\bibitem[\protect\citeauthoryear{Vieyro {\it et al.\/}}{2017}]{VRBR17}
	Vieyro, F. L., Romero, G. E., Bosch-Ramon, V. {\it et al.\/} 2017
	\aap\ 602, 64.
\bibitem[\protect\citeauthoryear{Wang {\it et al.\/}}{2018}]{W18} Wang, W.,
	Lu, J., Chen, X. \& Xu, R. 2018 SCIENCE CHINA Physics, Mechanics \&
	Astronomy in press.
\bibitem[\protect\citeauthoryear{Zhang}{2017}]{Z17} Zhang, B. 2017 \apjl\
	836, L32.
\bibitem[\protect\citeauthoryear{Zhang {\it et al.\/}}{2018}]{Z18} Zhang,
	Y.~G., Gajjar, V., Foster, G. {\it et al.\/} 2018 \apj\ 866, 149
	arXiv:1809.03043.
\end{thebibliography}
\end{document}